\documentstyle[psfig,referee]{mn}

\def\kms{\;{\rm km\,s^{-1}}}
\def\kmsmpc{\;{\rm km\,s^{-1}\,Mpc^{-1}}}
\def\mpcoh{\;h^{-1}\,{\rm Mpc}}
\def\hdens{\;h^3\,{\rm Mpc}^{-3}}

\begin{document}

\title[2dFGRS: rich galaxy clusters]{The 2dF Galaxy Redshift Survey: A 
targeted study of catalogued clusters of galaxies} 

\author[De Propris et al.]{
\parbox[t]{\textwidth}{
Roberto De Propris$^1$,
Warrick J. Couch$^1$,
Matthew Colless$^2$,
Gavin B. Dalton$^3$,
Chris Collins$^4$
Carlton M.\ Baugh$^5$,
Joss Bland-Hawthorn$^6$,
Terry Bridges$^6$,
Russell Cannon$^6$,
Shaun Cole$^5$,
Nicholas Cross$^7$,
Kathryn Deeley$^1$, 
Simon P.\ Driver$^7$,
George Efstathiou$^8$,
Richard S.\ Ellis$^9$,
Carlos S.\ Frenk$^5$,
Karl Glazebrook$^{10}$,
Carole Jackson$^2$,
Ofer Lahav$^{11}$,
Ian Lewis$^6$,
Stuart Lumsden$^{12}$,
Steve Maddox$^{13}$,
Darren Madgwick$^8$,
Stephen Moody$^{8,9}$
Peder Norberg$^5$,
John A.\ Peacock$^{14}$,
Will Percival$^{14}$,
Bruce A.\ Peterson$^2$,
Will Sutherland$^3$,
Keith Taylor$^9$}\\
\vspace*{6pt} \\
$^1$Department of Astrophysics, University of New South Wales, Sydney,
    NSW 2052, Australia; propris@bat.phys.unsw.edu.au \\
$^2$Research School of Astronomy \& Astrophysics, The Australian
    National University, Weston Creek, ACT 2611, Australia \\
$^3$Department of Physics, Keble Road, Oxford OX3RH, UK \\
$^4$Astrophysics Research Institute, Liverpool John Moores University,
    Twelve Quays House, Birkenhead, L14 1LD, UK \\
$^5$Department of Physics, South Road, Durham DH1 3LE, UK \\
$^6$Anglo-Australian Observatory, P.O.\ Box 296, Epping, NSW 2121,
    Australia\\
$^7$School of Physics and Astronomy, North Haugh, St Andrews, Fife,
    KY6 9SS, UK \\
$^8$Institute of Astronomy, University of Cambridge, Madingley Road,
    Cambridge CB3 0HA, UK \\
$^9$Department of Astronomy, Caltech, Pasadena, CA 91125, USA \\
$^{10}$Department of Physics \& Astronomy, Johns Hopkins University,
       Baltimore, MD 21218-2686, USA \\
$^{11}$Racah Institute of Physics, The Hebrew University, Jerusalem,
91904, Israel \\
$^{12}$Department of Physics, University of Leeds, Woodhouse Lane,
       Leeds, LS2 9JT, UK \\
$^{13}$School of Physics \& Astronomy, University of Nottingham,
       Nottingham NG7 2RD, UK \\
$^{14}$Institute for Astronomy, University of Edinburgh, Royal Observatory,
       Blackford Hill, Edinburgh EH9 3HJ, UK \\
}

\date{Received 0000; Accepted 0000}
\maketitle

\begin{abstract}

We have carried out a study of known clusters within the 2dF Galaxy
Redshift Survey (2dFGRS) observed areas and have identified 431 Abell,
173 APM and 343 EDCC clusters. Precise redshifts, velocity dispersions
and new centroids have been measured for the majority of these objects,
and this information has been used to study the completeness of these
catalogues, the level of contamination from foreground and background
structures along the cluster's line of sight, the space density of the
clusters as a function of redshift, and their velocity dispersion
distributions.  We find that the Abell and EDCC catalogues are
contaminated at the level of about 10\%, whereas the APM catalogue
suffers only 5\% contamination. If we use the original catalog
centroids, the level of contamination rises to approximately 15\% for
the Abell and EDCC catalogues, showing that the presence of foreground
and background groups may alter the richness of clusters in these
catalogues. There is a deficiency of clusters at $z \sim 0.05$ that may
correspond to a large underdensity in the Southern hemisphere.  From
the cumulative distribution of velocity dispersions for these clusters,
we derive an upper limit to the space density of $\sigma > 1000 \kms$
clusters of $3.6 \times 10^{-6} \hdens$. This result is used to
constrain models for structure formation; our data favour low-density
cosmologies, subject to the usual assumptions concerning the shape and
normalization of the power spectrum.

\end{abstract}
\section{Introduction.}

Rich clusters of galaxies are tracers of large-scale structure on the
highest density scales and therefore are important and conspicuous
`signposts' of its formation and evolution. While observational studies
of the structure and dynamics of rich clusters have by practical
necessity had to assume them to be isolated, spherically-symmetric
systems, recent massive N-body simulations of large-scale structure
growth (e.g. the {\small\it VIRGO\/} consortium; Colberg et al. 1998)
have shown a much more complex picture. Clusters are seen to be located
at the intersections of the intricate pattern of sheets, filaments and
voids that make up the galaxy distribution. They are formed through the
episodic accretion of smaller groups and clusters via collimated infall
along the filaments and walls (e.g. Dubinski 1998 and references
therein). As a result of this process, the large-scale structure that
surrounds the cluster gets imprinted upon it, both structurally (on
smaller scales) and dynamically.

Testing the predictions of the theoretical work, observationally, has 
not been easy since it requires large quantities of photometric and (in 
particular) spectroscopic data covering entire clusters and their 
surrounding regions. However, with the 2dF Galaxy Redshift Survey (2dFGRS;
Colless 1998, Maddox et al. 1998) -- the largest survey of its kind to be 
undertaken -- this problem can be addressed in a significant way. The
large ($\sim 10^{7} h^{-3}$ Mpc$^{3}$) and continuous volumes of space 
mapped by the survey together with its close to 1-in-1 sampling of the 
galaxy population, will ensure that it includes a large and representative 
collection of rich clusters, each of which is well sampled spatially over 
the desired large regions. Ultimately, when the survey is complete, it 
will be used in itself to generate a new 3D-selected catalogue of rich 
clusters, using automated and objective detection algorithms.  

The main purpose of this paper is to undertake a preliminary study of
catalogued clusters using these data and take a first look at such
issues as: the reality of 2D-selected clusters such as those in the
Abell catalogue, the incidence of serious projection effects and
contamination by foreground and background systems, the space density
of clusters and its variation as a function of redshift, richness and
cluster velocity dispersion. An additional by-product of the paper is
to present new redshift and velocity dispersion measurements for the
clusters, updating existing data in some cases and providing completely
new data in others. This will be used as the basis catalogue for an
analysis of: composite cluster galaxy luminosity functions and their
variation with cluster properties; spectrophotometric indices and their
dependence on local density; the star formation rates of galaxies in
clusters and their surroundings; the X-ray temperature -- velocity
dispersion relation, a study of bulk rotation in clusters and other
applications, which will be presented in separate papers. In addition,
this study will help define the nature of Abell clusters in 3D space,
so that objective cluster finding algorithms (to be applied to the 2dF
database upon completion of the survey) may be tailored to recover this
catalogue.

Our focus on the space density of clusters is motivated by the fact
that the {\it abundance\/} of clusters provides a probe of the
amplitude of the fluctuation power spectrum on characteristic scales of
approximately $10 \mpcoh$ -- corresponding to the typical cluster mass
of $\sim 5 \times 10^{14} h^{-1} {\rm M}_{\odot}$.  Once the average
density is determined, the cluster abundance can provide constraints on
the shape of the power spectrum. A well known example of this is the
observation that the Standard Cold Dark Matter (CDM) model, normalized
to match the cosmic microwave background anisotropies from the COBE
experiment, predicts an abundance of clusters in excess by one order of
magnitude over the observations.

The cluster mass function may therefore be exploited as a cosmological
test; however, determination of cluster masses is generally difficult.
For this reason, the distribution of velocity dispersions has often
been used as a surrogate (e.g. Crone \& Geller 1995). In particular,
the more massive, higher velocity dispersion clusters, are less likely
to suffer from biases and incompleteness, and their space density may
provide constraints on models for the formation of large scale
structure.  Previous work indicates that clusters with $\sigma > 1000
\kms$ are relatively rare (e.g. Mazure et al. 1996 and references
therein). Depending on the normalization and shape of the fluctuation
spectrum, this can be used to constrain cosmological parameters. In
most common models, the rarity of these objects is taken to imply a
low value of the matter density. 

The plan of the paper is as follows: In Section 2, we give a brief
overview of the 2dFGRS observations. Section 3 then describes the
selection of clusters for this study and how the members in each were
identified using the 2dFGRS data; we derive redshifts and velocity
dispersions for a sample of objects with adequate data.  In section 4
we address the issues of contamination of the cluster catalogues and
selection of appropriate samples for comparison with theoretical
models. This is followed in Section 5 by a determination of the space
density of the different sets of catalogued clusters studied here, and
then in Section 6 we analyse this quantity as a function of cluster
velocity dispersion, comparing it with cosmological models.  Finally, a
summary of our results is given in Section 7. A cosmology with
$H_0=100\,h \kmsmpc$ and $\Omega_{0}=1$ is adopted throughout this
paper.

\section{Observations}

The observational parameters of the 2dFGRS are described in detail
elsewhere (Colless et al. 2001) and so only a brief summary is given
here:  The primary goal of the 2dFGRS is to obtain redshifts for a
sample of 250\,000 galaxies contained within two continuous strips (one
in the northern- and the other in the southern-galactic cap regions)
and 100 random fields, totalling $\sim 2000$\,deg$^{2}$ in area, down
to an extinction-corrected magnitude limit of $b_{J}=19.45$.  The input
catalogue for the survey is based on the APM catalogue published by
Maddox et al. (1990a,b), with modifications as described  in Maddox et
al. (2001, in preparation).

Observations are carried out at the 3.9\,m Anglo-Australian Telescope
(AAT), using the Two-degree Field (2dF) spectrograph, a fibre-fed
instrument capable of obtaining spectra for 400 objects simultaneously
over a two-degree field (diameter). The instrument is described in
Lewis et al. (2001, in preparation). For the 2dFGRS, 300\,line/mm
gratings blazed in the blue are used, yielding a resolution of $\sim
9$\,\AA\ FWHM and a wavelength range of 3500-7500\AA. To date, the
observing efficiency, accounting for weather losses and instrument
down-time, has averaged $\sim 50$\%, with the overall redshift
completeness running at $\sim 95$\%, based on a typical exposure time
(per field) of 3600\,s.  The spectra are all pipeline reduced at the
telescope, with redshifts being measured using a cross-correlation
method and subject to visual verification in which a quality index $Q$,
which ranges between 1 (unreliable) and 5 (of highest quality), is
assigned to each measurement. As of July, 2001, we had collected
195,\,497 unique redshifts, including 173,\,084 galaxies with good
quality spectra (the sample used here). The balance of objects consists
of galaxies with poor spectra and stars misclassified as galaxies.

\section{Cluster Selection and Detection}

\subsection{The cluster catalogues}

Clusters for our study were sourced from the catalogues of Abell (Abell
1958; Abell, Corwin \& Olowin 1989, hereafter ACO), the APM (Dalton et al.
1997) and EDCC (Lumsden et al. 1992).

Abell and collaborators selected clusters from visual scans of Palomar
Observatory Sky Survey red plates and from SERC-J plates. For each
cluster, a counting radius was assigned, equivalent to $1.5 \mpcoh$
(the Abell radius), adopting a redshift based on the magnitude of the
10$^{th}$ brightest galaxy ($m_{10}$). The number of cluster galaxies
between $m_3$ and $m_3$+2, where $m_3$ is the magnitude of the third
brightest galaxy, was then used to assign a richness parameter, after
subtracting an estimate for background and foreground contamination.
Abell (1958) used a local background from areas of each plate with no
obvious clusters, whereas ACO employed a universal background derived
from integration of the local luminosity function.

Both the APM and EDCC use machine-based magnitude-limited galaxy
catalogues from the UK Schmidt plates. A full description of the APM
selection algorithm is given by Dalton et al. (1997). The APM cluster
survey used an optimized variant of Abell's selection algorithm which
uses a smaller radius to identify clusters and a richness estimate
which is coupled to the apparent distance to compensate for the effects
described by Scott (1956). This produces richness and distance
estimates for the APM clusters which are found to be robust, and which
give well-defined estimates of the completeness limits for the
catalogue. The large-scale properties of the final 2-D catalogue are
found to be consistent with the observed 3-D distribution (Dalton et
al. 1992).

Lumsden et al. (1992) adopt an approach similar to Abell; they bin
their data in cells and lightly smooth the distribution to identify
peaks, using a procedure akin to that of Shectman (1985). EDCC clusters
are then related to the Abell catalogue, with the catalogue listing a
richness class and magnitudes for the first, third and tenth ranked
galaxies.

By nature of its visual selection, the Abell catalogue is somewhat
subjective, and prone to contamination from plate-to-plate variations
and chance superpositions. Lucey (1983) and Katgert et al. (1996)
estimate that about 10\% of the clusters with richness class $R\geq 1$
suffer from contamination, whereas Sutherland (1988) argues for a
15--30\% level of contamination over the entire sample, including the
poorer clusters. Here contamination is defined as the presence of
foreground or background structure that substantially boosts the
apparent richness of the system, in some cases allowing the inclusion
in the catalogue of objects that would not satisfy the minimum richness
criterion.  This definition is, of course, somewhat arbitrary and
subjective; we adopt a somewhat more quantitative definition when we
examine the issue of contamination later in the paper. Sutherland \&
Efstathiou (1991) also infer the presence of significant spurious
clustering in the Abell catalogues due to completeness variations
between plates, although they do not quantify this further.

Both the APM and the EDCC claim to be more complete than the Abell
catalogue, especially for poor clusters, and to be less affected by
superposition and contamination. The EDCC claims to be complete for all
clusters within the context of the stated selection criteria; EDCC is
built to imitate the Abell catalogue and a comparison shows that about
50\% of the clusters are in common between the two catalogues. The APM
uses a smaller counting radius than Abell and is claimed to be more
complete for poorer clusters and to be more objectively selected
(Dalton et al. 1997).

\subsection{Cluster identification and measurement}

We searched the 2dFGRS catalogue for clusters whose centroid, as given
in the above catalogues, lay within 1 degree of the centre of one of
the observed survey tiles. In doing so, our policy was to consider {\it
all\/} clusters in each of the three catalogues without any
pre-selection based on richness or distance class or any other
property.  The Abell catalogue is, in theory, limited to clusters with
$z < 0.2$; however, it includes clusters with estimated redshifts that
are substantially higher (e.g. Abell 2444 in the sample being
considered here). Although theses objects may well be too distant for 
2dFGRS to detect, they are included in our Tables nonetheless, since it is 
generally difficult to estimate cluster redshifts {\it a priori}. 

If the centroid of a catalogued cluster was found in one of the 2dFGRS
tiles, we then searched the 2dFGRS redshift catalogue for objects
within a specified search radius of the cluster centroid.  The search
radius used was that particular to the catalogue from which the cluster
originated. This isolates a cone in redshift space containing putative
cluster members along with foreground and background galaxies. We then
inspected the Palomar Observatory Sky Survey (POSS) plates for the
brightest cluster galaxy: in most cases this was a typical central
cluster elliptical with optical morphology consistent with a brightest
cluster galaxy and could therefore be easily identified as the cluster
centre. Where this was not possible, in some clusters, we adopted the
brightest cluster member with an image consistent with early-type
morphology. We repeated our search procedure to produce more accurate
lists of candidate members.

An important consideration in this context, is the adaptive tiling
strategy used in 2dF observations (Colless et al. 2001).  Here,
complete coverage of the survey regions is achieved through a variable
overlapping (in the Right Ascension direction) of the 2dF tiles.  In
the direction of rich clusters where the surface density of galaxies is
high, more overlap is clearly required. Hence we have to tolerate some
level of incompleteness in the peripheries of our fields at this stage
of the survey; this is a {\it temporary} situation, the implications of
which will be discussed later in this section. In Table 1 we quote the
completeness, viz. the fraction of 2dFGRS input catalogue objects
within our search radius whose redshifts have been measured for each
cluster field.

\subsection{Cone Diagrams}

This transformation of the projected 2D distribution of galaxies upon the
sky (and which the identification of a cluster was based) into a 3D
one, presented us with three general cases as far as cluster visibility
was concerned: (i)\,The cluster was easily recognizable as a distinct
and concentrated collection of galaxies along the line of sight with no
ambiguity at all in its identification. Cone diagrams for three such
examples (A0930, A3880 and S0333) are shown in Fig. 1.  (ii)\,Several
concentrations of galaxies were found along the line of sight. Where
one was particularly dominant, then cluster identification was
generally unambiguous, but foreground and background contamination was
clearly significant. Two such examples (A1308, A2778) are shown in
Fig.  2. If the different concentrations were of similar richness then
cluster identification became ambiguous and required further analysis
via our redshift histograms (see below).  An example of such a case
(S0084) is also shown in Fig. 2.  (iii)\,There were no clearly defined
concentrations of galaxies at all within the cone and the cluster, at
this stage, could not be identified. Three such examples (A2794, A2919,
S1129) are shown in Fig. 3, where the `cluster' appears in redshift
space to be a collection of unrelated structures. Note that the opening
angles of the cone diagrams are far larger than the search radius,
corresponding to a metric radius of 6 Mpc for the adopted cosmology;
this is done in order to show both the cluster and its surrounding
large-scale structure. In contrast, the redshift histograms that we now
discuss have been constructed from objects just within the search
radius, in order to facilitate identification of the cluster peak.

To consolidate and quantify our cluster identifications, redshift
histograms of the galaxies within the Abell radius were constructed and
examined.  These are also included in Figs 1--3 for each of the cone
diagrams that are plotted. For the ambiguous case (ii) types, where the
redshift histogram contained multiple and no singly dominant peaks (see
A2778 in Fig. 2), the peak closest to the estimated redshift of the
cluster was taken to be our identification. In none of the case (iii)
situations did the redshift distribution allow us to identify a
significant peak. All peaks that were found in the direction of each
cluster are listed in Table 1. Notes indicate the presence
of fore/back-ground systems.

\subsection{Redshifts and Velocity Dispersions}

Mean redshifts and velocity dispersions were calculated from the
redshift distributions, not only for the identified clusters but also
for all the other significant peaks seen. In doing so, we followed the
approach of Zabludoff, Huchra \& Geller (1990; ZHG) to identify and
isolate cluster members. The basis of this method is that (as shown in
the redshift histograms in Figures 1--3) the contrast between the
clusters and the fore/back-ground galaxies is quite sharp. Therefore,
physical systems can be identified on the basis of compactness or
isolation in redshift space, i.e., on the gaps between the systems: in
this latter case, if two adjacent galaxies in the velocity distribution
are to belong to the same group, their velocity difference should not
exceed a certain value, the {\it velocity gap}. ZHG use a two-step
scheme along these lines, in which first a fixed gap is applied to
define the main system and then a gap equal to the velocity dispersion
of the system is applied to eliminate outlying galaxies. The choice of
the initial gap depends somewhat on the sampling of the redshift
survey: e.g., ZHG use a 2000 $\kms$ gap. In order to avoid merging well
separated systems into larger units (as we are better sampled than ZHG)
we adopt a 1000 $\kms$ gap.  The choice of 1000km/s was found to be
optimum in that it (i) avoids merging sub-cluster systems into a large
and spurious single system, and (ii) is large enough to avoid
fracturing real systems into many smaller groups. We also note that the
value of 1000 km/s that was used, is consistent with previous work and
such a value is borne out by the distribution of velocity separations
in the cluster line of sight pencil beams (cf., Katgert et al. 1996)
and is operationally simpler than implementing a friends-of-friends
algorithm. In principle this choice may introduce a bias with redshift,
as the luminosity functions are less well sampled for more distant
clusters, and this is the reason why most of our analysis below is
carried out on the nearer portion of the sample.

The redshift bounds of the `peak' corresponding to the cluster were set
by proceeding out into the tails on each side of the peak centre until
a velocity separation between individual galaxies of more than $1000
\kms$ was encountered.  In other words, we define the cluster peak as
the set of objects confined by a $1000 \kms$ void on either side in
velocity space. The peak can have any width in velocity space, but is
required to be isolated in redshift space. We then calculated a mean
redshift and velocity dispersion for the galaxies in the peak and
ranked them in order of redshift separation from the mean value. We
next identified the first object on either side of the mean whose
separation in velocity from its neighbour (closest to the mean)
exceeded the velocity dispersion, and then excised all objects further
out in the wings of the distribution. The mean redshift and velocity
dispersion were then recalculated following the prescription of Danese,
de Zotti \& di Tullio (1980), which provide a rigorous method to estimate
mean redshifts, velocity dispersions and their errors based on the
assumption that galaxy velocities are distributed according to a
Gaussian.

If the final, excised sample contained fewer than 10 objects, a
velocity dispersion was not calculated, since values based on such
small numbers are too unreliable (Girardi et al.  1993). We quote the
standard deviation of the mean in place of the velocity dispersion but
make no use of it in our analysis. These clusters are, however,
included in our tables below, and in our analysis of completeness and
the space density of clusters presented in the following sections. By
imposing this number threshold, we should also decrease our sensitivity
to sampling variations (due to the increased fibre collisions in denser
fields and therefore lower completeness for cluster fields).

This procedure is a simplified form of the `gapping' algorithm
suggested by Beers, Flynn \& Gebhardt (1990). Previous work has
generally employed the pessimistic 3$\sigma$ clipping technique of
Yahil \& Vidal (1977).  One advantage is that the ZHG technique 
does not assume a Gaussian distribution of velocities and discriminates
against closely spaced peaks, corresponding to a lower $\sigma$ clip in
the case of a pure normal distribution. On the other hand, the
3$\sigma$ clipping method is more effective at removing spurious high
velocity dispersion objects when the fields are sparsely sampled. A
comparison between the two methods has been carried out by Zabludoff et al
al. (1993): while the results are usually consistent within the
1$\sigma$ error, there is a tendency for 3$\sigma$ clipping to yield
somewhat lower velocity dispersions.

\begin{figure*} 
{\psfig{file=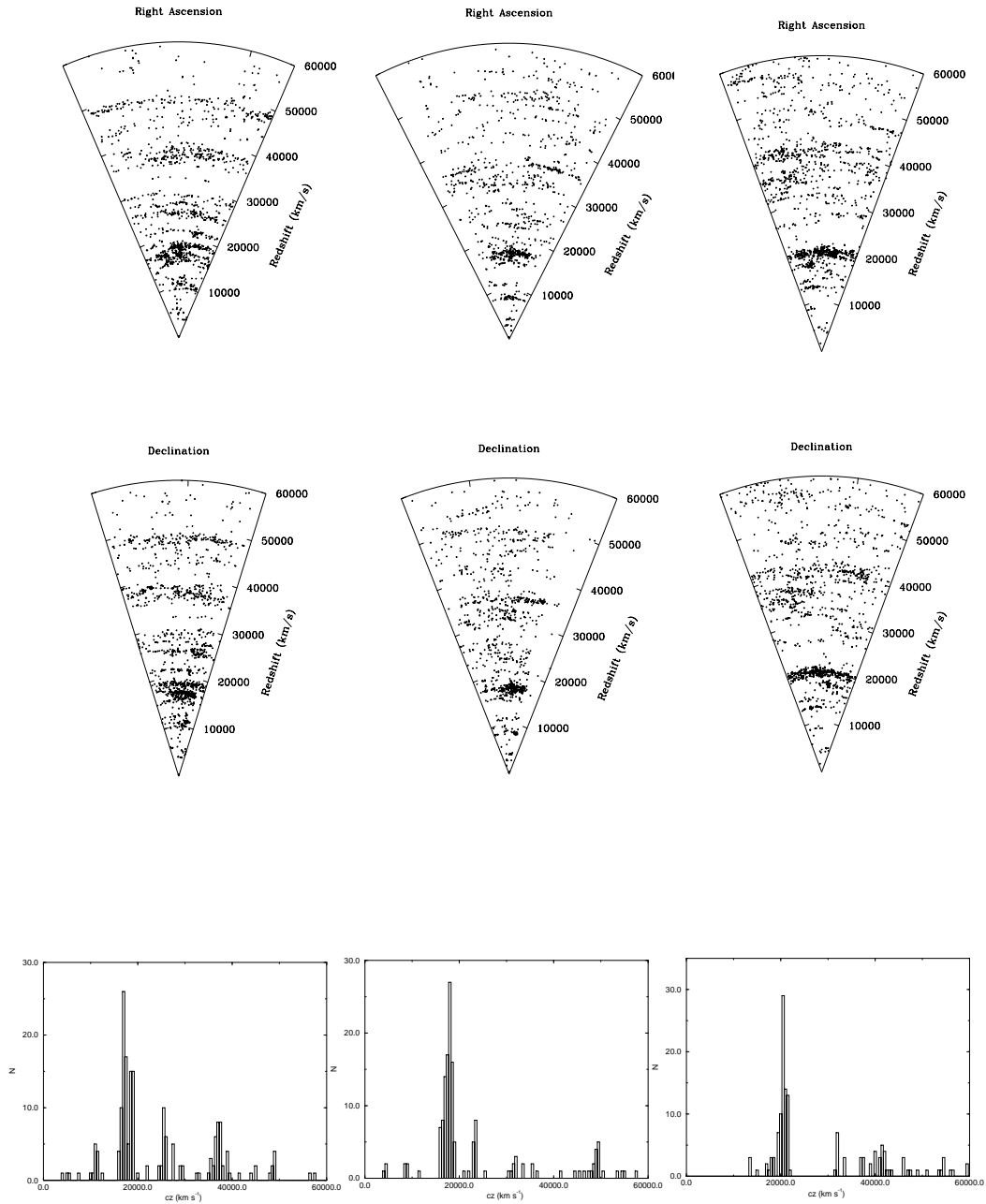,height=8.5in}} 
\caption{Cone
diagrams and redshift histograms for the fields centred on Abell 0930,
3880 and S0333 (from left). The aperture is a circular one with radius
corresponding to 6 Mpc at the cluster redshift.} \end{figure*}

\begin{figure*}
{\psfig{file=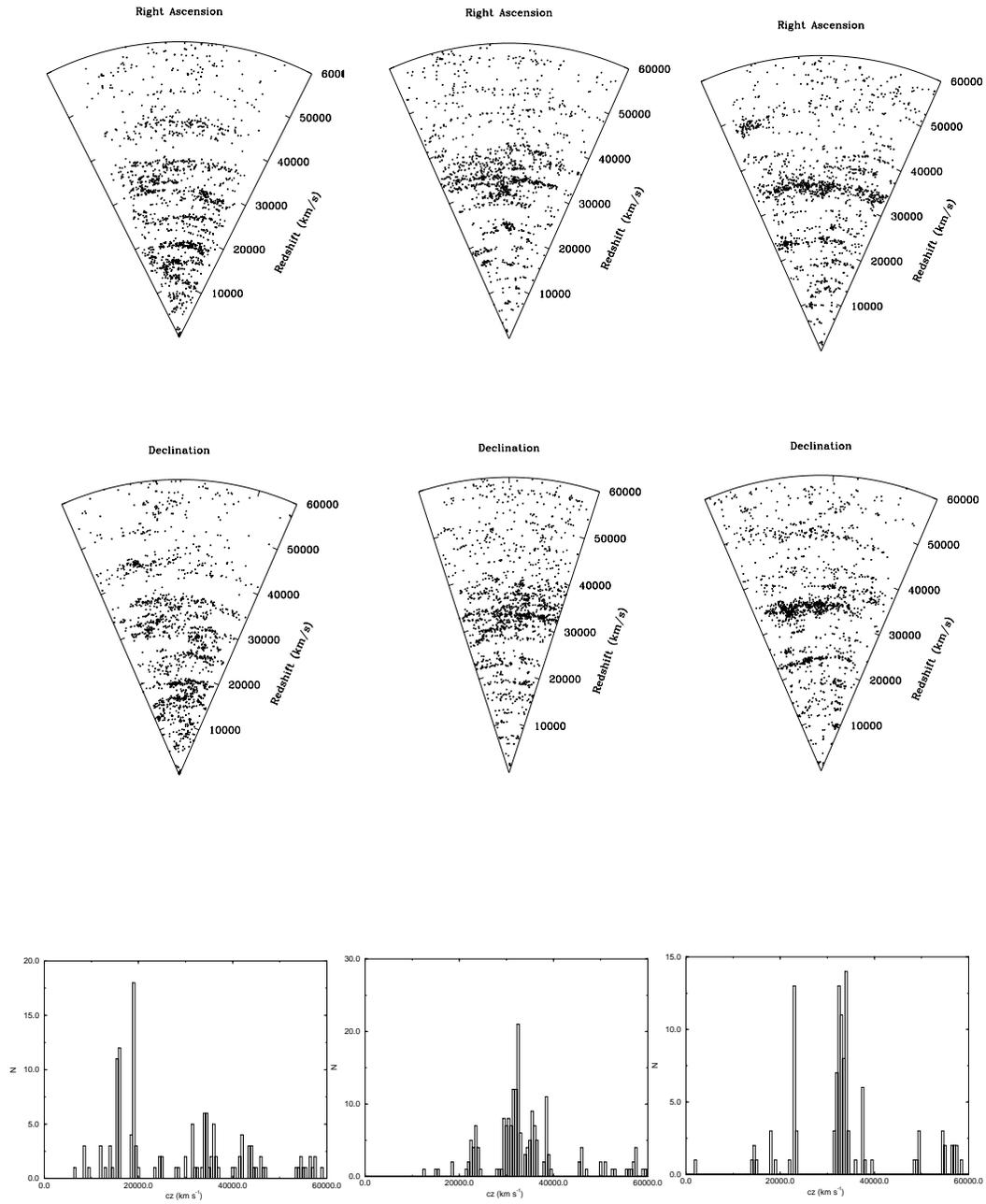,height=8.5in}} 
\caption{Cone diagrams and redshift histograms for the fields centred 
on Abell 1308, 2778 and S0084} \end{figure*}

\begin{figure*} 
{\psfig{file=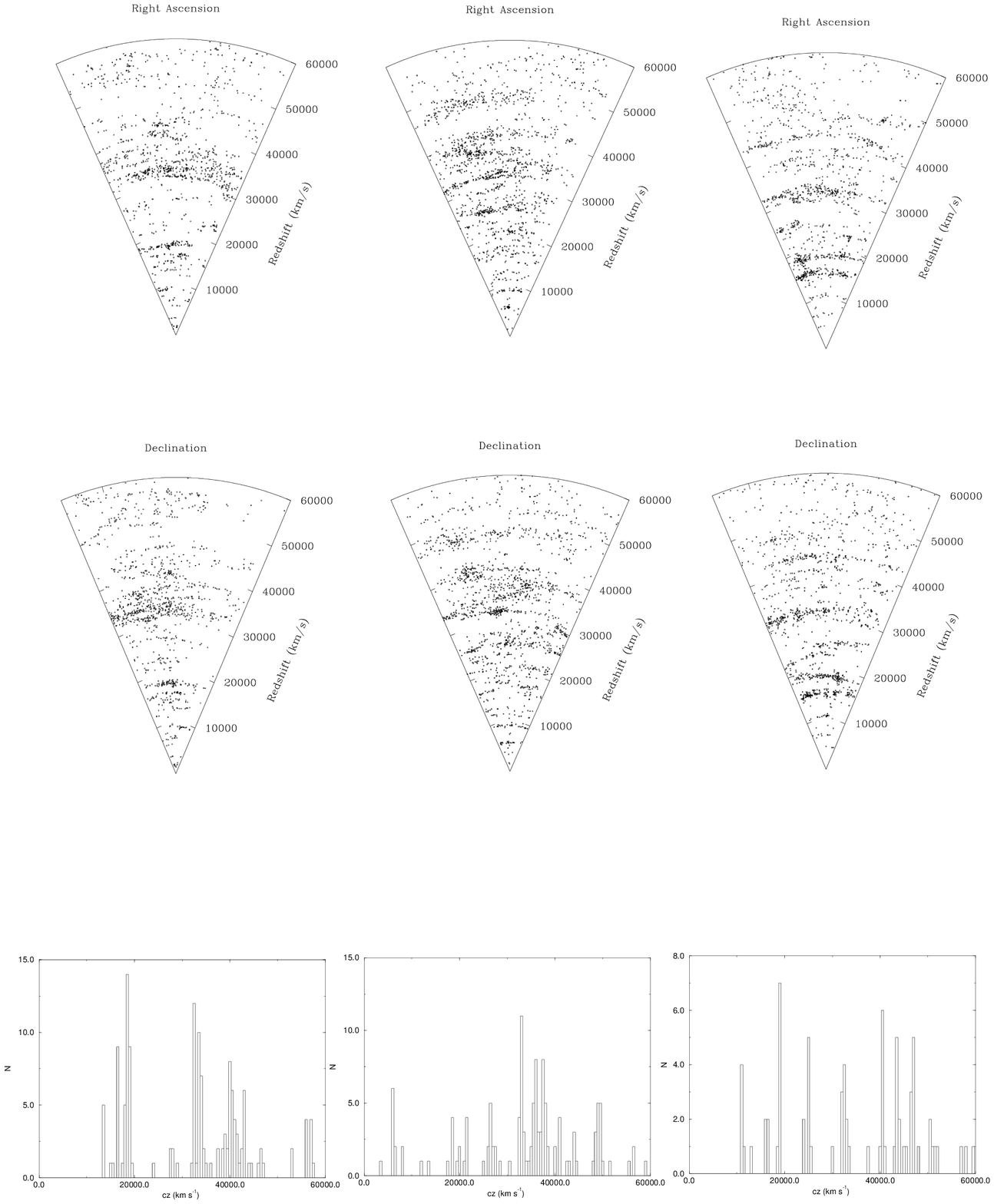,height=8.5in}} 
\caption{Cone diagrams and redshift histograms for the fields centred 
on Abell 2794, 2919 and S1129} \end{figure*}

\subsection{The Cluster Tables}

Table 1 lists all unique clusters detected (where unique means detected
in a single catalogue, avoiding counting objects more than once if they
are present in more than one catalog: the order of preference is Abell,
APM and EDCC). This includes 1149 objects (including double or triple
systems where more than one identifiable cluster or group is present in
the line of sight) and 753 single clusters (i.e. assuming only one of
the eventual multiple systems corresponds to the catalogued cluster).
Of these 413 are in the Abell/ACO catalogues, 173 in APMCC and 343 in
EDCC.  The structure of the table is as follows: column 1 is the
identification, columns 2 and 3 are cross-identifications in other
catalogues, columns 4 and 5 the RA and Dec of the cluster centroid (see
above), column 6 the redshift we derive along with its error, column 7
the velocity dispersion, column 8 the number of cluster members, and
column 9 the redshift completeness (expressed as a percentage) in the 2
degree (diameter) tile the cluster is located.  Column 10 contains
essential notes. Literature data are from the recent compilations of
Collins et al. (1995), Dalton et al.  (1997) and Struble \& Rood
(1999), unless noted. The first few lines of the table are printed
here: the entire table is available in ASCII format from {\tt
http://bat.phys.unsw.edu.au/~propris/clutab.txt}

\begin{figure*}
\setcounter{figure}{0}
{\psfig{file=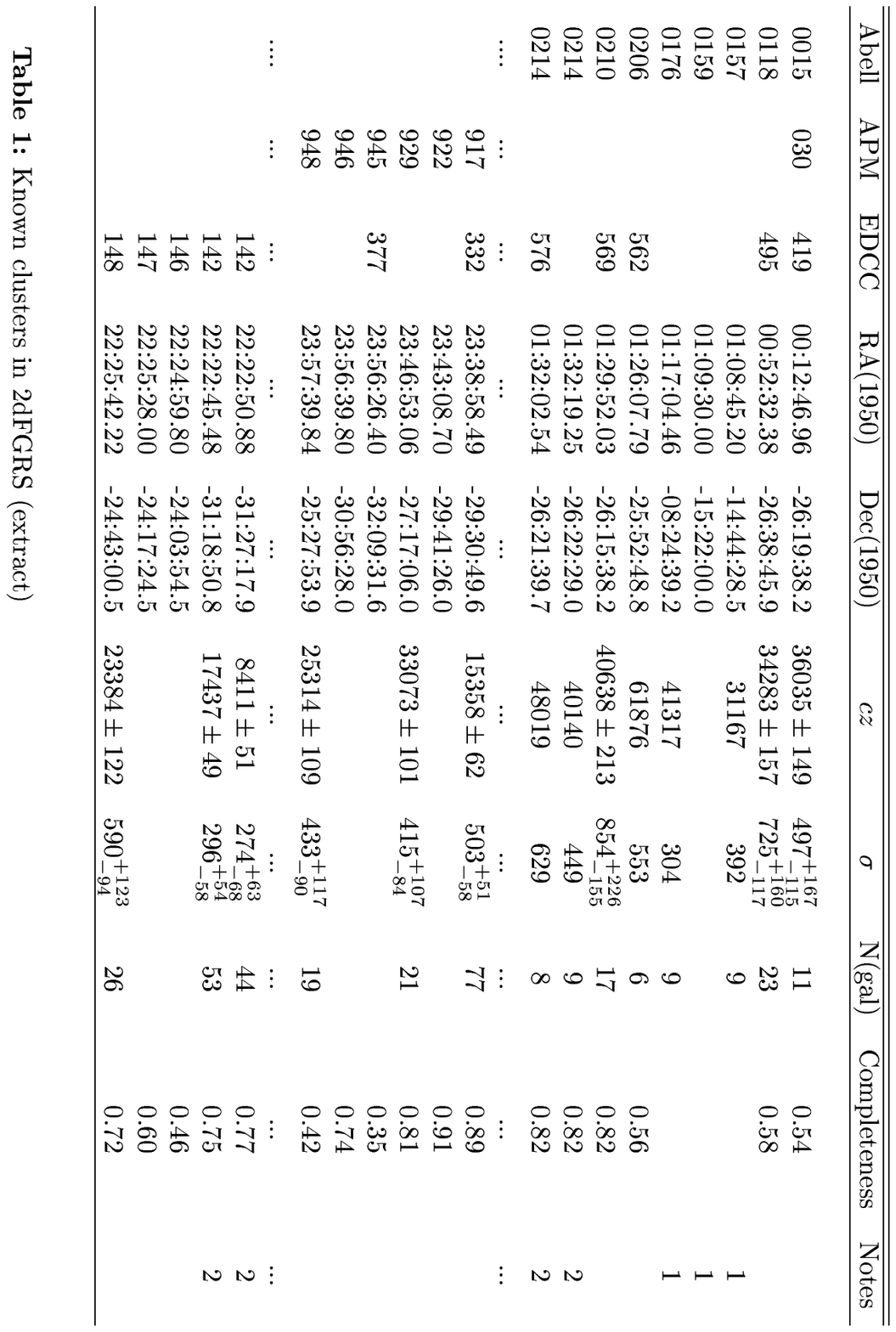,height=8.5in}}
\end{figure*}

Having assembled the cluster redshifts, measured both here using the
2dFGRS data and previously by other workers, we can compare the two to
provide an external check on our new 2dFGRS values. We compare
redshifts for clusters which have more than 6 measured members in
2dFGRS. To avoid confusion, we only consider clusters with a single
prominent peak, since in the cases where more than one structure is
present in the beam, the identification with the cluster is ambiguous.
This comparison is shown graphically in Fig. 4 where we see a good
one-to-one relationship between the two. Formally we find a mean
difference between ours and other redshift measurements of $\Delta cz =
89 \pm 307 \kms$. This excludes a small number of objects where the 2dF
and literature redshift disagree by large values: such cases appear to
occur when the cluster centroid in the original catalogue is
misidentified or when only one or two galaxies are used to derive
the previously published redshift.

\begin{figure*}
\setcounter{figure}{3}
{\psfig{file=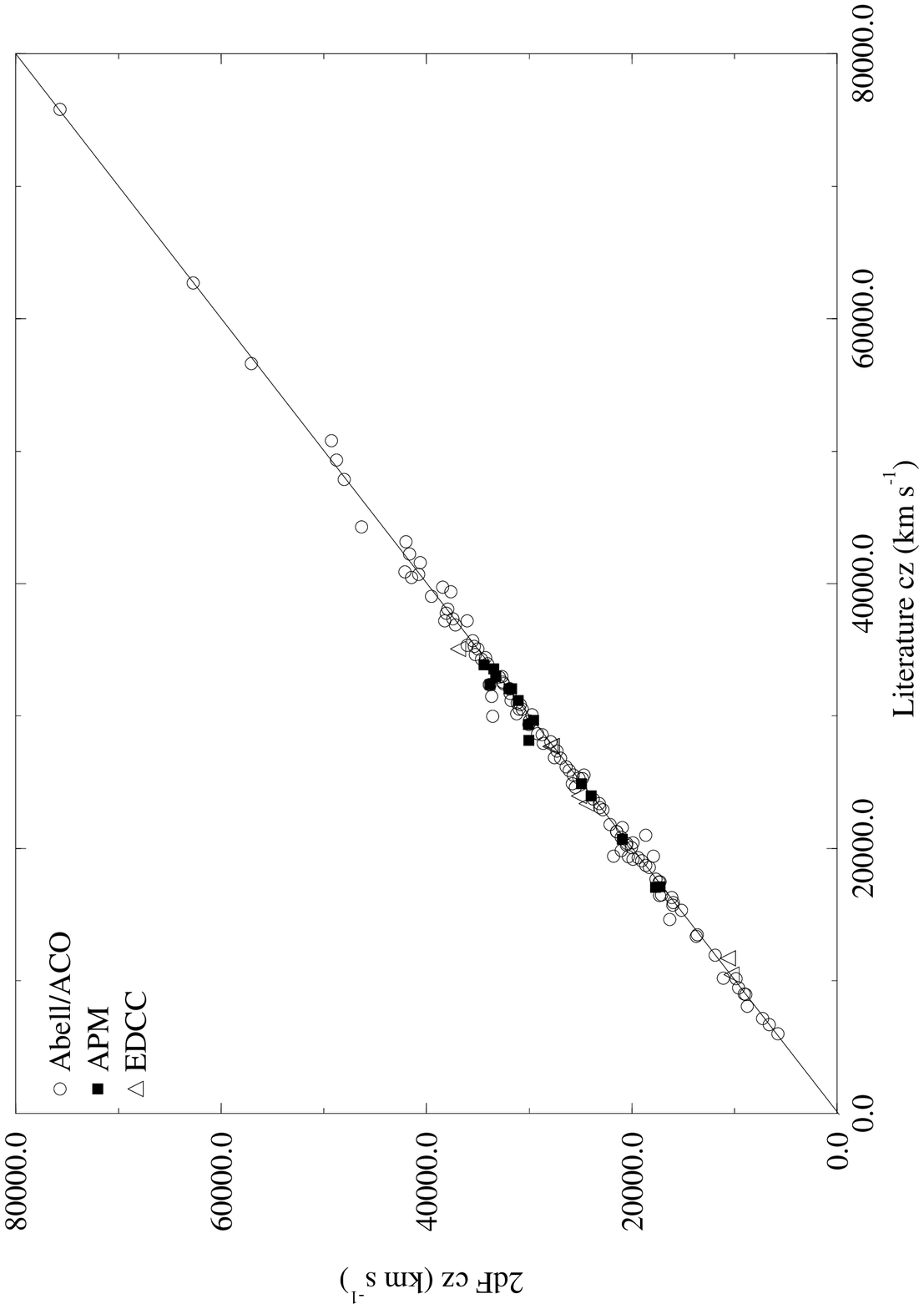,height=8.5in}}
\caption{Comparison between literature and 2dF cluster redshifts}
\end{figure*}

Finally, in Table 2 we summarize the total numbers of clusters from
each catalogue found within the 2dFGRS. It is important to stress that
the sum of these totals does not represent the number of unique
clusters that are studied here, since there is some overlap between the
3 different cluster catalogues (although we have analysed them
separately according to the definitions of each catalogue -- see
above). We show the level of overlap by listing alongside the totals
for each catalogue -- in column 2 of Table 2-- the numbers of these
clusters that are also found in the other 2 catalogues.

About one third (32\%) of all Abell clusters are identified with
an EDCC cluster and 10\% with an APM cluster. Conversely, 24\% of APM 
clusters have an Abell and 29\% an EDCC counterpart. For EDCC, 39\% of 
clusters are also identified in Abell and 15\% in the APM. Note that this 
comparison is confined to just the southern strip and does not include any 
of the clusters in the original Abell (1958) catalogue.

\begin{table*}
\setcounter{table}{1}
\centering
\begin{minipage}{140mm}
\caption{Summary of Cluster Identifications}
\begin{tabular}{cccc}
\hline
Catalogue & N(clusters) & N(Redshifts) & N($\sigma$)  \\
Abell   &   413 (42 APM, 133 EDCC) & 263 & 208 \\
APM     &   173 (42 Abell, 50 EDCC) & 84 & 75 \\
EDCC    &   343 (133 Abell, 50 APM) & 224 & 174 \\
\hline
\hline
\end{tabular}
\end{minipage}
\end{table*}

\section {Cluster completeness and contamination}

Important to any quantitative analysis based on the clusters found here
is the need to identify volume-limited sub-samples, underpinned by a
good understanding of the completeness of the input cluster catalogues
and how the derived velocity dispersions maybe biased with redshift and
cluster richness. We note in this regard that a properly selected 3D
sample will be derived using automated group finding algorithms once
the survey has reached its full complement of galaxies and the window
function is more regular.

In order to derive estimates of completeness and contamination and
normalize the space density of clusters to determine the distribution
of velocity dispersions (Section 5 below), we need to define properly
volume--limited samples and correct our observations for incompleteness
deriving from the adopted window function and detection efficiency.
Here we adopt two routes: the standard approach has been to define
`cuts' in estimated redshift space to derive a (roughly) volume limited
sample, adopting a richness limit to insure that the sample will be
reasonably complete. We first comment on the accuracy of estimated
redshifts and any empirical relation that exists between estimated and
true (2dFGRS) redshift; afterwards we use this relation and our
redshifts together to determine an estimate for the space density of
clusters and choose an adequately complete sample. We also adopt a more
simplistic approach, determining the space density of all clusters for
which we have redshifts. Although this sample is incomplete, by
definition, it is strictly volume limited (also by definition) and
provides a useful lower limit to the quantities of interest.

Previous studies which have targeted clusters from available 2D
catalogues, have approached this problem by using appropriate cuts in
richness and $m_{10}$. For example, the ENACS survey (Katgert et al.
1996) studied all $R > 1$ Abell clusters with m$_{10} < 16.9$. This
sample is approximately volume limited to $z \sim 0.1$, but incomplete
in that it does not include all clusters with $z < 0.1$. Estimated
redshifts have also been used to derive information on cosmology from
analysis of the distribution of Abell clusters (e.g. Postman et al.
1985). It is therefore of interest to consider the accuracy of
photometric redshift estimators via comparison with our more accurately
determined 2dFGRS spectroscopic values.

\subsection{The Abell/ACO Sample}

Figure 5(a) plots estimated redshift (using the formulae in Scaramella
et al. 1991) vs. 2dF redshift for the Abell sample. We see an acceptable
linear relationship, with some tendency to saturate at very high redshifts
(where the estimated redshift is slightly higher than the measured one).

Figure 5(b) shows what fraction of the catalogued clusters are in each of 
the different estimated redshift bins (width $\Delta z=0.02$), plus the
fractional distributions for both those clusters identified in 2dFGRS and 
those that were missed. We see that we are reasonably complete to a redshift 
of about $\sim 0.10$ and our completeness drops beyond that as cluster 
galaxies drop below the survey magnitude limit. 

We split our sample at $z = 0.15$ where approximately equal numbers of
objects are missed or identified and plot the distribution of cluster
richnesses (as measured from m$_3+2$).

For objects with $z_{est} < 0.15$, the distributions of richnesses for
identified and missed objects are similar [Figure 5(c)]. Surprisingly,
this is also true for objects with $z_{est} > 0.15$ in Figure 5(d). The
fraction of missed objects in the $z < 0.15$ group rises rapidly in the
last two redshift bins. The similar richness distributions suggest that
at least some of the missed objects are really spurious
superpositions.  We also plot the fractions of recovered and missed
clusters as a function of completeness in each tile of 2dFGRS in Figure
6: we see no strong trend. We also divide the sample according to
richness, at the median richness of the sample ($R=50$). Although there
is a small tendency for poorer clusters to be missed in low 2dF
completeness regions (as one would expect), we find no strong trend in
this sense. This suggests that we would be able to find the clusters,
if they are real.  We calculate that about 25\% of clusters in the
$z_{est} < 0.15$ group are missed, which would be consistent with the
estimate (van Haarlem, Frenk \& White 1997) that about one third of all
Abell clusters are actually superpositions of numerous small groups
along the line of sight.

\begin{figure*}
{\psfig{file=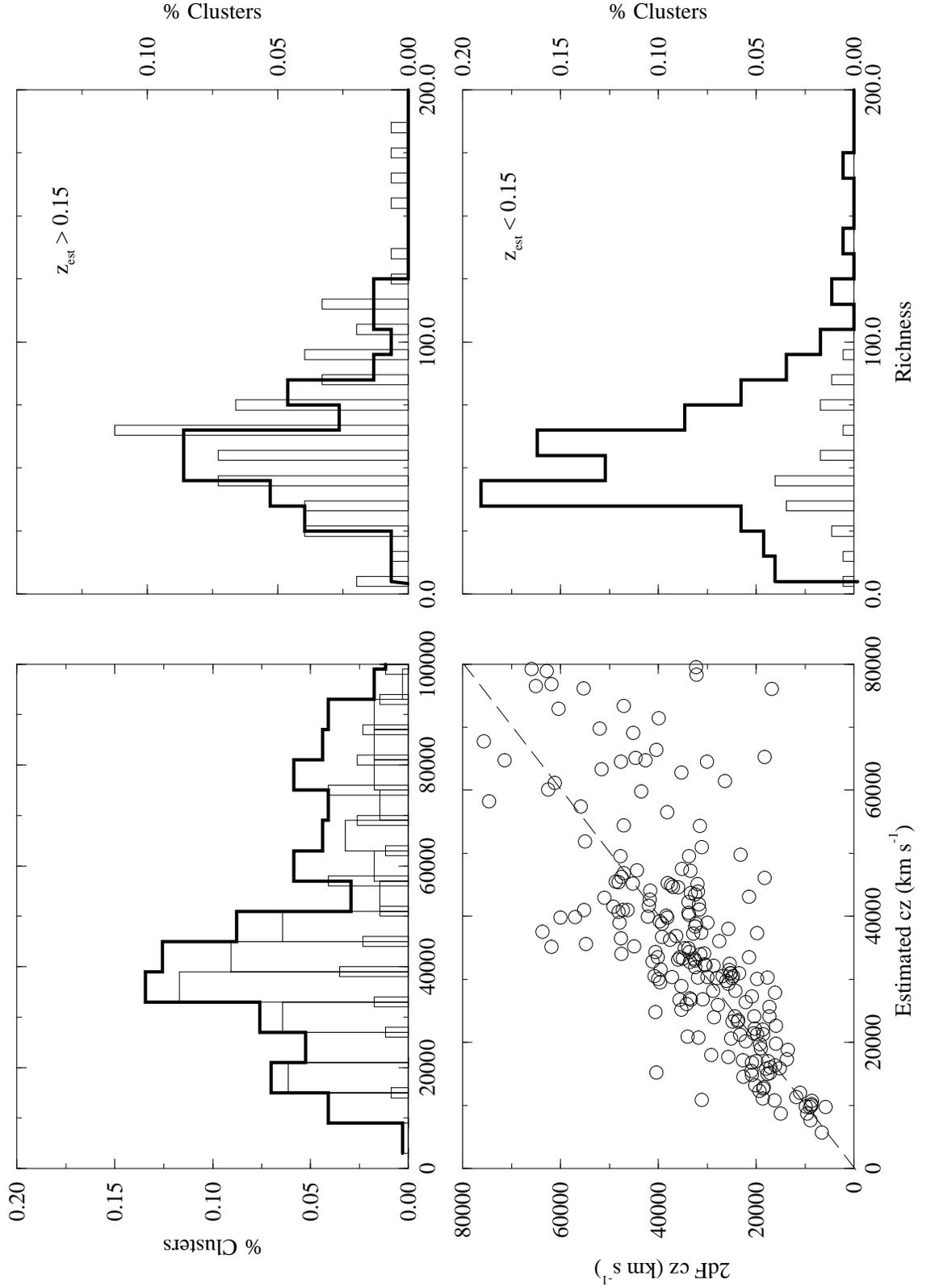,height=8.5in}}
\caption{Data for the Abell sample. Panel (a) compares estimated and measured
redshifts; panel (b) shows the fraction of clusters as a function of
estimated redshift: the broad thin-lined histogram represents the catalogued
clusters, the thick-lined histogram represents the clusters identified within
2dFGRS, and the thin-lined narrow bars represent clusters that were missed.
Panels (c) and (d): as for panel (b), but the fractions are plotted as 
a function of richness for the $z_{est} < 0.15$ and $z_{est} > 0.15$ samples, 
respectively;  here the thick-lined histogram represents the detected 
clusters while the thin-lined histogram represents the missed clusters.
}
\end{figure*}

\begin{figure*}
{\psfig{file=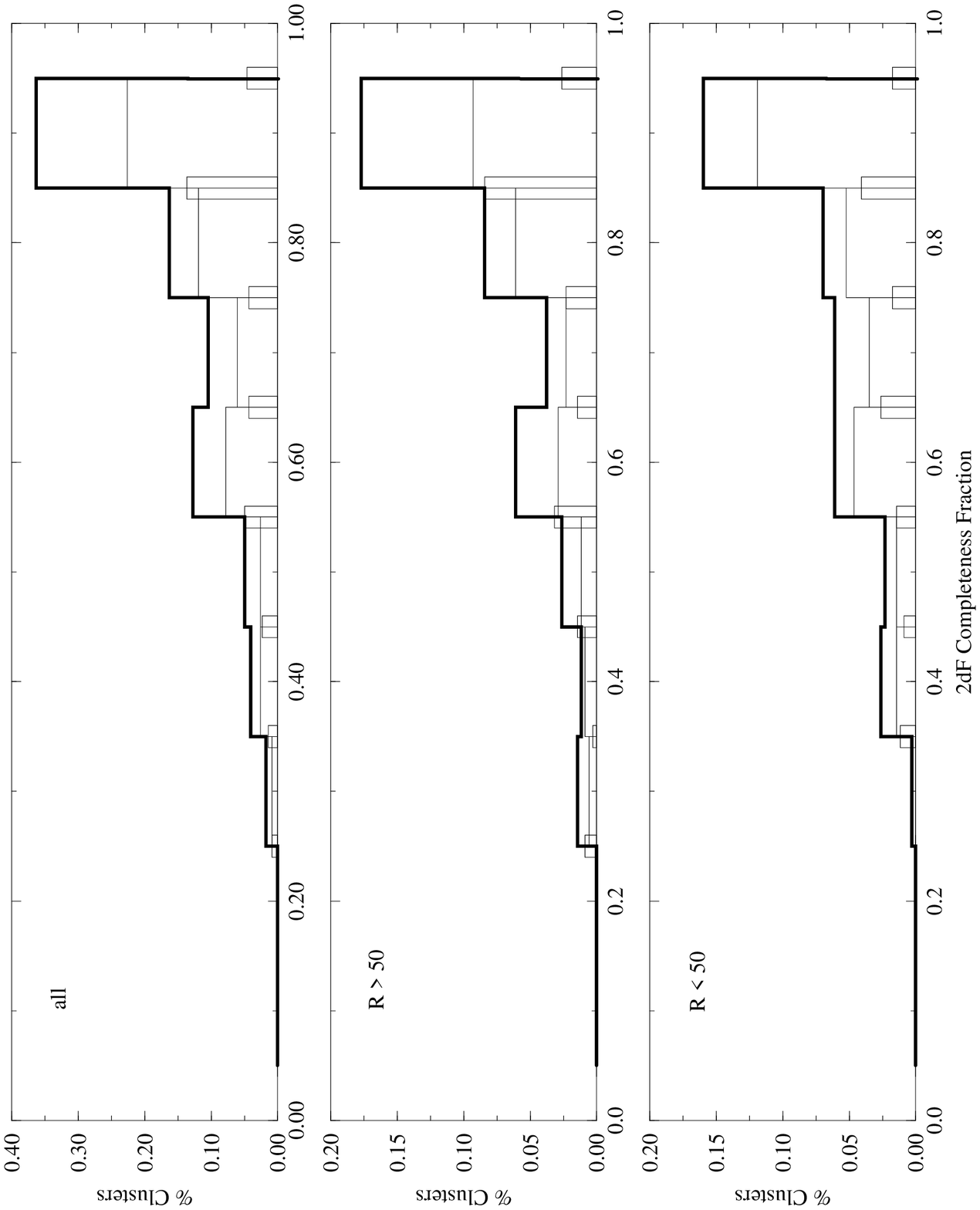,height=8.5in}}
\caption{The same fractions as plotted in Fig. 5(b-d) for the sample of
Abell clusters, but here plotted as a function of the redshift completeness
in the 2dFGRS tile in which the cluster is located. The broad thin-lined
histogram represents the catalogued clusters, the thick-lined histogram
represents the clusters identified within 2dFGRS, and the thin-lined narrow bars
represent clusters that were missed. We plot all clusters in the upper panel,
those with $R > 50$ in the middle and those with $R < 50$ in the lower panel.}
\end{figure*}

\subsection{The APM Sample}

We plot the estimated vs. measured redshift for the APM sample in
Figure 7(a). The relationship is reasonably linear but the APM
estimated redshifts saturate at $z \sim 0.12$. This effect derives from
the magnitude limit used in the parent galaxy catalogues, where
star-galaxy separation becomes unreliable at $b_j \sim 20.5$. Figure 7(b)
shows the 2dFGRS detection success rate as a function of completeness
in the 2dFGRS tile: we note that there is some tendency for APM clusters 
to be missed at low completeness. We plot the fractions of all catalogued 
clusters, those found and those missed for the APM in Figure 7(c)
and we see that whereas the sample is complete to $z < 0.07$, clusters
are increasingly missed at higher redshifts. The distribution of 
richnesses [Figure 7(d)] shows that most of the missed objects tend to be 
the poorer systems, as one would expect. The more homogeneous behaviour 
of the APM cluster catalogue (in terms of completeness as a function of
redshift and richness) is probably a reflection of the more objective 
search algorithm used (cf. Abell's).

\begin{figure*}
{\psfig{file=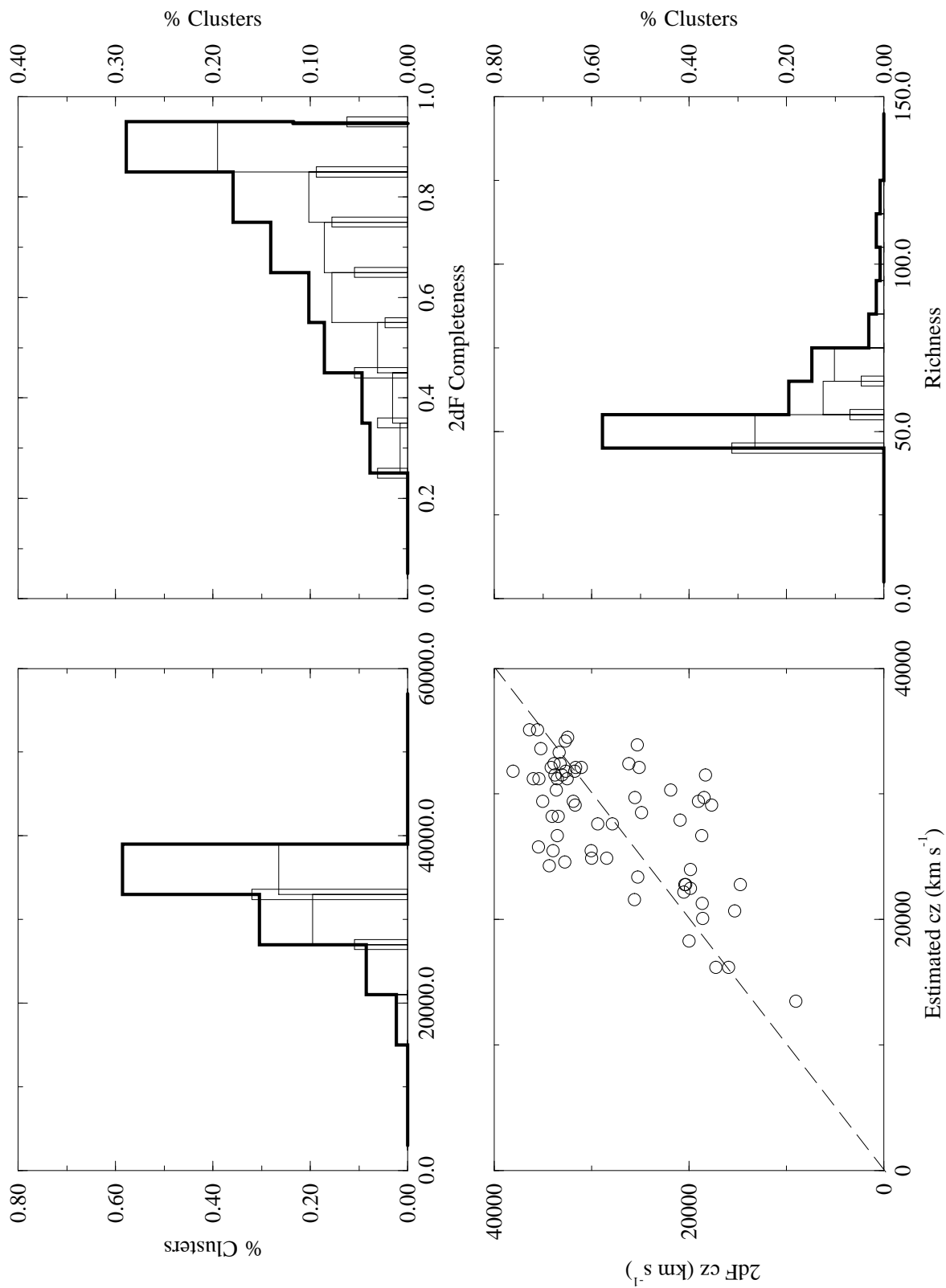,height=8.5in}}
\caption{Data for the APM sample. Panel (a) compares estimated and measured
redshifts; panel (b) shows the fraction of clusters as a function of
estimated redshift: the broad thin-lined histogram represents the
catalogued clusters, the thick-lined histogram represents the clusters
identified within 2dFGRS, and the thin-lined narrow bars represent
clusters that were missed.  Panel (c): as for panel (b), but the
fractions are plotted as a function of richness; here the thick-lined
histogram represents the detected clusters while the thin-lined
histogram represents the missed clusters.  Panel (d): as for panel (b)
but the fractions are plotted as a function of completeness in each
tile.
}
\end{figure*}

\subsection{The EDCC Sample}

We plot estimated vs. measured redshifts for the EDCC sample in Figure
8(a). Here we see that EDCC tends to systematically overestimate the 
cluster redshift. We tried to derive a more accurate formula for EDCC 
estimated redshifts based on the formalism of Scaramella et al. However, 
we see that the $m_{10}$ indicator for EDCC saturates quickly and we are 
unable to determine a more accurate relation between estimated and true 
redshift. The distribution of completeness fractions in tiles for 
catalogued, recovered, and missed objects are shown in panel (b) where we 
see a trend for clusters to be missed in low completeness regions (as one 
would expect). Panel (c) shows the distributions as a function of 
estimated redshift: here we find little difference between the three 
classes of clusters. Panel (d) shows the richnesses: again, recovered and 
missed objects follow the same distributions. 

\begin{figure*}
{\psfig{file=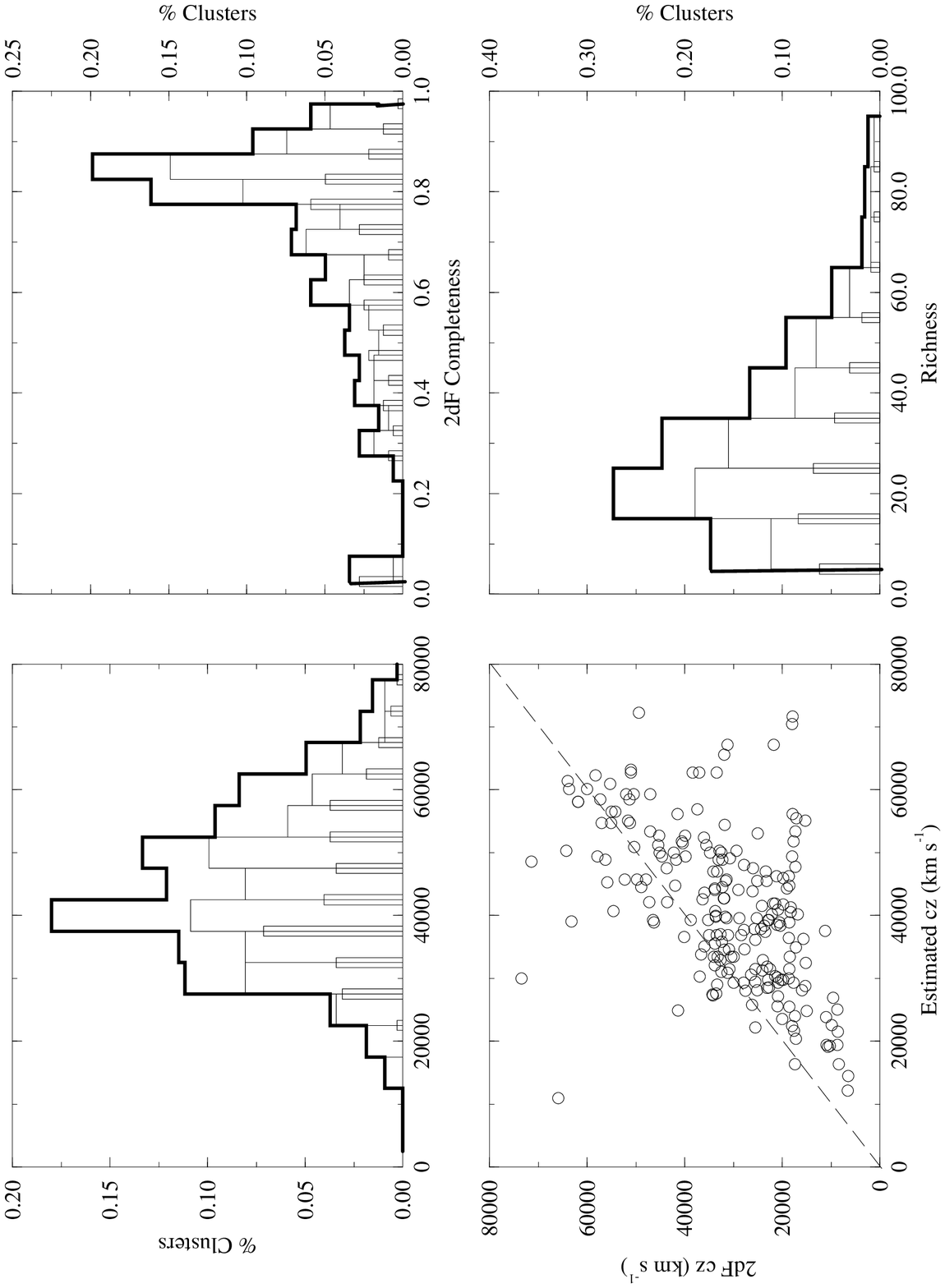,height=8.5in}}
\caption{As per Figure 6 for the EDCC clusters.}
\end{figure*}

\subsection{Contamination of Cluster Catalogs}

The broad relation that exists between estimated and true $cz$ has been
used in previous studies to define an estimated $cz$ such that, given
the spread in the relation, the sample will be approximately
volume--limited within a specified $cz$, although it will not
necessarily be complete. We now go through this exercise here, choosing
limits rather conservatively in order to minimise the level of
incompleteness. By way of example, we derive `volume limited' cuts from
estimated redshifts below and determine the level of contamination: we
also use these relationships in the next section where we consider the
space density of clusters.

For the Abell sample we choose a limit of $z < 0.11$, where we are
reasonably complete. This includes 110 clusters with 100 redshifts.  Of
these 9 have significant foreground or background structure. Here and
for the other clusters as well, we define ``significant'' to mean that
we were able to derive at least a redshift and in some cases a velocity
dispersion for the background or foreground systems (these are
tabulated in Table 1 as well). About 10\% of Abell clusters are therefore
contaminated systems by our definition. If we use the original centroids
we obtain a contaminated fraction of 15\%. This is due to the fact that
fore/back-ground groups shift the real cluster centre away from
its proper position.

For the APM catalogue we use the entire sample. Of the 173 clusters, only
5 are contaminated by fore/back-ground groups, i.e about 3\%. A slightly
higher fraction (5\%) is derived from the original centroids. This
lower fraction is simply due to the smaller radius used by APM, which 
increases the contrast between cluster and field.

The EDCC is more complicated, as the relationship between estimated
and true redshifts is non-linear and shows a sizeable offset. We choose
an estimated $cz$ of $50000 \kms$ to include all objects within $30000
\kms$. This includes 234 clusters, with 165 redshifts. By our definition
15 of these objects show contamination, equivalent to 8\%, similar to 
the Abell sample. If we adopt the original centres we find a level of
about 13\% contamination. This is well within the estimate by Collins
et al. and is not peculiar to the EDCC catalog but rather an unavoidable
consequence of the selection procedure imitating Abell's.

We therefore confirm the earlier studies by Lucey (1983) and Sutherland
(1988) that the Abell catalogue suffers from contamination at
approximately the 15\% level, if the original cluster centres are
used.  The EDCC catalogue behaves similarly. The APM seems to be best
at selecting real clusters; this is most likely due to the smaller
search radius employed by Dalton et al. (1992) and the higher richness
cut used to produce the APM catalogue. If we use more accurate centres
the level of contamination is reduced, suggesting that in some cases
the position and richness of the clusters are shifted by the presence of the 
fore/back-ground group.

\section{The Space Density of Clusters}

We have used the 2dFGRS to select clusters over a wide range of
richness and to establish a more accurate volume-limited sample than
possible from photometric indicators. Having done so, we now examine
the space density of clusters as a function of redshift in each of the
catalogues, in order to choose a redshift within which the sample is at
least reasonably complete. The density of clusters as a
function of redshift within $0.01$ intervals is shown in
Fig. 9. We also plot a corresponding sample from the RASS1 survey of De
Grandi et al.  (1999). The RASS1 is an X-ray selected survey of Abell
clusters spanning about one third of the Southern sky: for this reason
the sample is only semi-independent from ours, although it does not
fully overlap with our slices. Since the true space density of clusters is
expected to be approximately constant over this range of redshifts, the
observed general decline in the cluster space density at $z\geq 0.1$
must reflect the incompletenesss of the Abell, APM and EDCC
catalogues at these limits (plus our own inability to detect clusters
as some complex function of richness, distance and incompleteness). 

Within $z < 0.15$ (chosen as the redshift range in which we are nearly
complete), we see in Fig. 9 that there is considerable fluctuation of
the space density. Furthermore, the Abell et al. and EDCC clusters both
exhibit a density minimum at $z\sim 0.05$ (as also seen in the galaxy
distribution;  Cross et al.  2000) at approximately the 2$\sigma$
level.  The deficit extends across the entire Southern strip of the
survey and possibly beyond, corresponding to a 200 Mpc $h^{-1}$ scale
void. While this is potentially very interesting, we must be extremely
cautious at this stage that this is not just a sampling effect that
results from the small (and hence unrepresentative) volume so far
covered by the 2dFGRS at these low redshifts. We note that a similar
effect has been noted by Zucca et al. (1997) in the ESO slice survey
(ESP), and can be explained in the same manner if one considers the
location of the ESP within the APM Galaxy Survey map. A comparison with
the wider RASS1 survey of X-ray selected clusters also plotted in Fig.
7, shows no evidence of such a structure.

However, three semi-independent samples show this feature at
statistically significant levels. It would be difficult to devise a
selection effect working against $z \sim 0.05$ clusters (only) in a 2D
sample. Subject to the caveats above, these data are suggestive of a
large underdensity in the Southern hemisphere, in the direction sampled
by the APM. This would account for the low normalization of the bright
APM counts without requiring strong evolution at low redshift (Maddox
et al.  1990c) and for the differences in the amplitude of the ESP and
Loveday et al. (1992) field luminosity functions. This deficit is not
seen in some other surveys because of the Shapley concentration, which
masks the underdensity centered close to the South Galactic pole. For
instance, the REFLEX survey reports an overdensity at this redshift
which is attributed to the Shapley structure (Schuecker et al. 2001).

\begin{figure*}
{\psfig{file=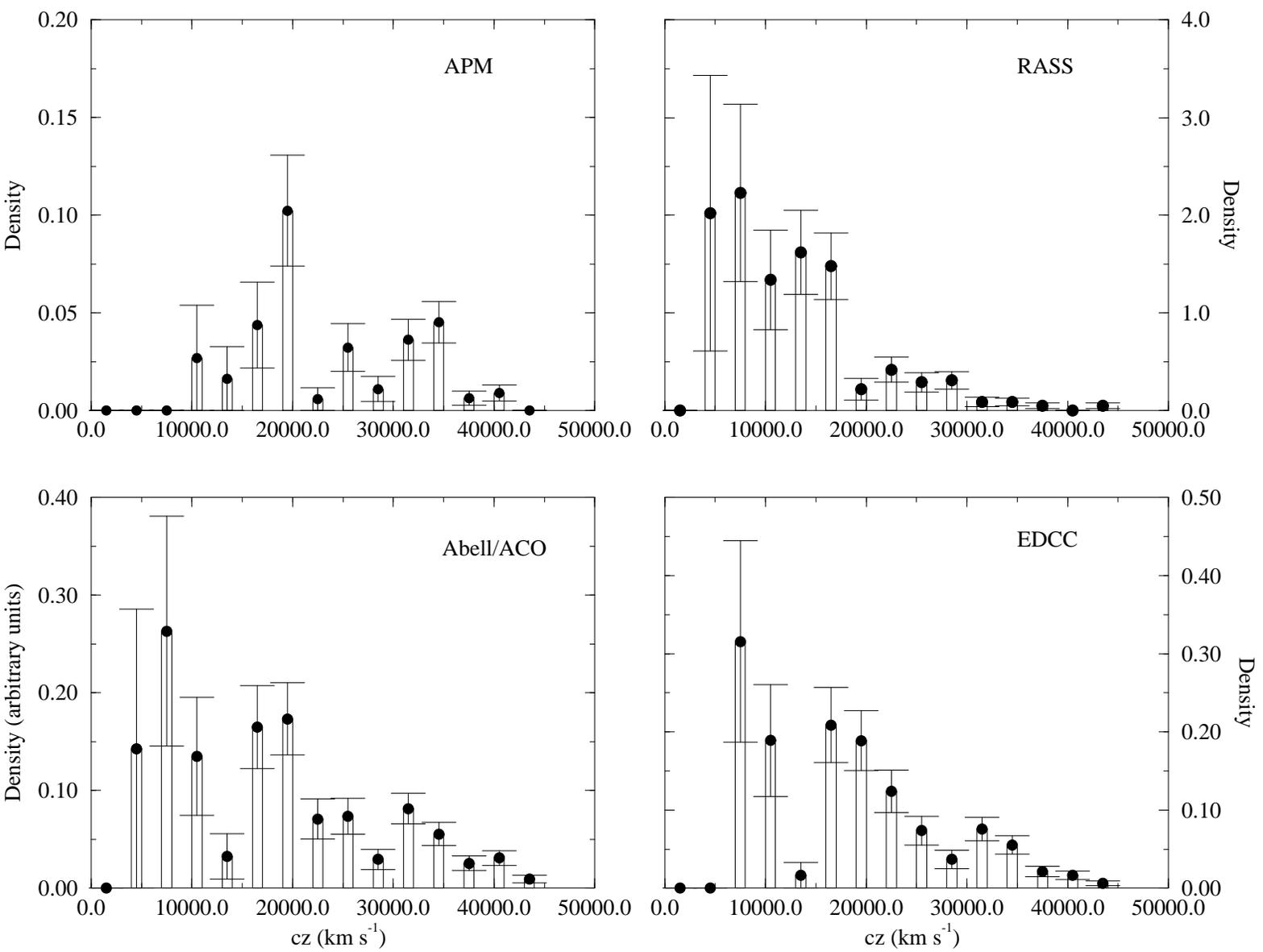,height=8.5in}}
\caption{Variation of space density (normalized to volume) for Abell,
APM and EDCC clusters and the RASS1 sample. Units of density are
arbitrary.}
\end{figure*}

In order to derive the distribution of cluster velocity dispersions to
be discussed in the next section, we need to determine the {\it true}
space density of catalogued Abell clusters. Naturally, this is but a
lower limit to the space density of {\it all} clusters, that can only
be derived from a 3D selected sample, but, at least for the richer
clusters, our sample should be complete. We restrict our attention to
Abell clusters, which are the most commonly used sample of objects.

As we have seen, it is possible to use the linear relationship between
estimated and true redshift for the Abell sample to define a reasonably
complete sample to $z \sim 0.11$. In the two survey strips we have 
surveyed a total of 984.8 square degrees. We therefore derive a space
density of ($27.8 \pm 2.8) \times 10^{-6} \hdens$ for all Abell clusters, 
and ($9.0 \pm 1.7) \times 10^{-6} \hdens$ for clusters of richness class
1 or greater. In comparison, Scaramella et al. derive a space density
of about $6 \times 10^{-6} \hdens$ and Mazure et al. (1996; ENACS) obtain 
$8.6 \times 10^{-6}$. Our result is in good agreement with the ENACS
value but somewhat higher than that of Scaramella et al.

\section{Velocity Dispersion Distribution}

The cumulative distribution of velocity dispersions provides
constraints on cosmological models of structure formation, via the
shape of the power spectrum of fluctuations. The power spectrum at
large scales can be determined from the COBE data (and subsequent cosmic
microwave background experiments), whereas cluster mass functions yield
limits on small scales. Although it is generally difficult to estimate
cluster masses, the distribution of velocity dispersions may be used as
a substitute. In particular, the space density of the most massive
(high $\sigma$) clusters, is a good discriminant between theoretical
models.

We assume that the distribution of velocity dispersions for clusters
with $z < 0.11$ represents the underlying true distribution. Some
support for this is given by Fig. 10, where we plot velocity dispersion
vs.  redshift and find no obvious correlation. This suggests that our
sample is `fair' in the sense that we are not systematically losing
clusters at any particular velocity dispersion.

\begin{figure*}
{\psfig{file=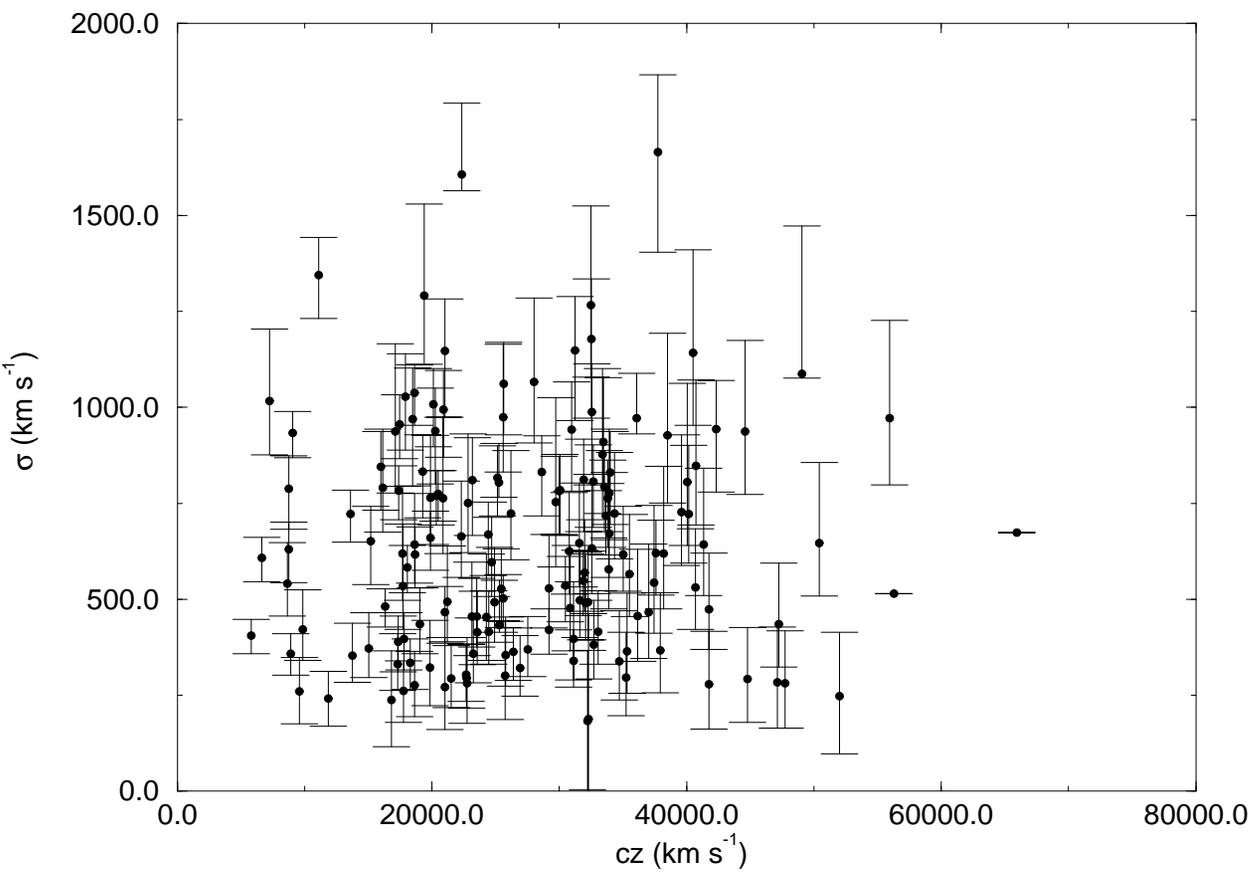,height=8.5in}}
\caption{Derived velocity dispersion vs. redshift for the Abell sample,
showing lack of correlation}
\end{figure*}

We plot our data in Fig. 11 (filled circles), together with previous
work by Zabludoff et al. (1993), Girardi et al.  (1993) and Mazure et
al. (1996) (all as lines). For the sake of comparison, we renormalize
these data to our local density. These should be taken with some
caution, especially at low velocity dispersions, where our sample
includes low richness objects (and all the samples become incomplete at
some level), but should be reasonable at high velocity dispersions,
where our results are in acceptable agreement with previous data.

The most robust result of our analysis is the confirmation of a
relative lack of high $\sigma$ clusters. As a matter of fact, since
interloper galaxies cause a spurious high $\sigma$ tail in the
distribution (van Haarlem et al., 1997), we feel we can derive
a significant value to the space density of N($\sigma > 1000
\kms$) clusters. We consider only clusters whose {\it derived}
redshifts place them within $z < 0.11$. This is equivalent to: 
$3.6 \pm 1 \times 10^{-6} \hdens$ and may be compared with
theoretical models by Borgani et al (1997), for instance: our data are
in good agreement with a Cold+Hot Dark Matter model; a $\Lambda$CDM
model with $\Omega_M=0.3$ underpredicts the space density of clusters
whereas one with $\Omega_M=0.5$ slightly overpredicts it; $\tau$CDM
models are acceptable as long as $\sigma_8 < 0.67$; open CDM models
with $\Omega_M=0.6$ are in good agreement with our results and Standard
CDM models normalized to COBE (as are all models in Borgani et al.) are
inconsistent with our derived space density. The data therefore favour
low matter densities or small values of $\sigma_8$ (where $\sigma_8$ is
the rms fluctuation within a top-hat sphere of $8 \mpcoh$ radius).
This would bring cluster results in better agreement with the COBE data
(e.g. Bond \& Jaffe 1999).

\begin{figure*}
{\psfig{file=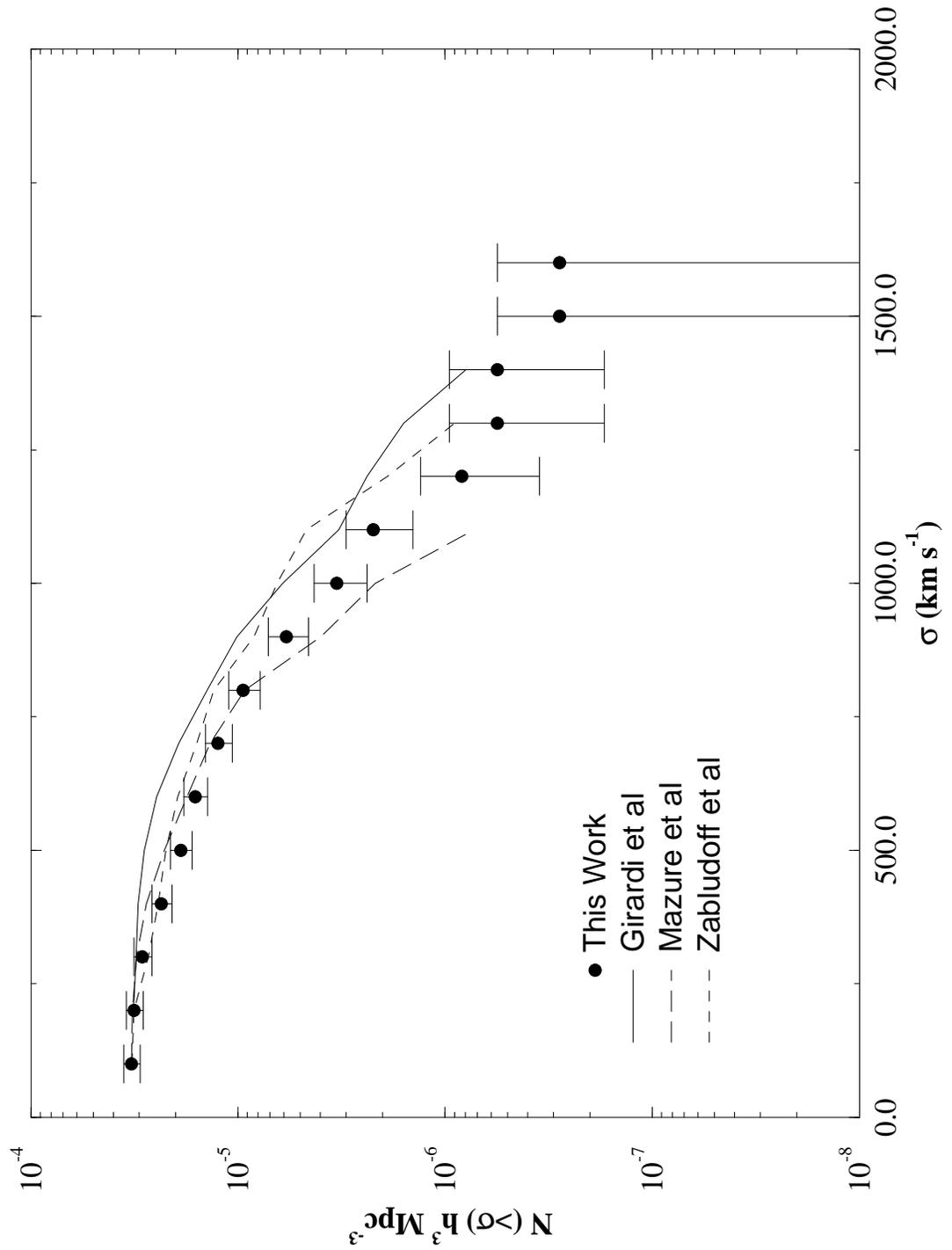,height=8.5in}}
\caption{Distribution of velocity dispersions for our sample and
previous work. We have renormalized Girardi et al and Mazure et al
data for the sake of comparison.}
\end{figure*}

\section{Summary}

We have analyzed a sample of 1149 previously catalogued clusters of
galaxies that lie within the 2dFGRS. The results of this analysis can be 
sujmarised as follows:

\begin{itemize}

\item New redshifts (and velocity dispersions) have been derived for a 
sample of 263 (208) clusters in the Abell sample, 84 (75) APM clusters and 
224 (174) EDCC clusters.

\item Of the 1149 clusters, 753 appear to have no counterpart in each of
the other catalogues and are thus unique.

\item The level of contamination of our clusters by fore/back-ground
groups is about 10\% for the Abell sample. However, if we select on
the original centroids, we confirm the earlier results of Lucey (1983) and Sutherland (1988) that for about 15--20\% of the Abell and EDCC clusters, background and foreground groups substantially boost the derived surface 
density and may lead to poor groups being erroneously identified as clusters.
This shows that the presence of interloper groups and galaxies may
skew the apparent richness and structure of clusters.

\item The space density of rich Abell clusters is broadly consistent
with previous work. For all Abell clusters the derived space density is
($27.8 \pm 2.8) \times 10^{-6} \hdens$; for $R>1$ clusters, we find a 
space density of ($9.0 \pm 1.7) \times 10^{-6} \hdens$. This is broadly consistent with, but better determined than, previous work.

\item We find evidence for the existence of an underdensity of clusters in 
the southern hemisphere at $z \sim 0.05$.

\item We derive an upper limit to the space density of clusters with velocity
dispersion greater than $1000 \kms$.  This is shown to be inconsistent
with some models of structure formation and to favour generally low
matter densities and low values of the $\sigma_8$ parameter.

\end{itemize}

\def\ref{\par\noindent\hangindent\parindent}

\section{Acknowledgements}

R.D.P. and W.J.C. acknowledge funding from the Australian Research Council 
for this work. We are indebted to the staff of the Anglo-Australian 
Observatory for their tireless efforts and assistance in supporting 
2dF throughout the course of the survey. We are also grateful to the
Australian and UK time assignment committees for their continued support
for this project.

\section{References}

\ref Abell G. O., 1958, ApJS, 3, 211
\ref Abell G. O., Corwin, H. G. \& Olowin, R. 1989, ApJS, 70, 1
\ref Beers T. C., Flynn K., \& Gebhardt K., 1990, AJ, 100, 32
\ref Bond J. R., \& Jaffe A. H., 1999, Phil. Trans. Roy. Soc. 357, 57
\ref Borgani S., Gardini A., Girardi M., \& Gottl\"ober S., 1997, 
New Astronomy, 2, 199
\ref Colberg J. M., et al. 1998 in {\it The Evolving Universe: Selected Topics
on Large Scale Structure and on the Properties of Galaxies}, Astrophysics
and Space Science Library, vol 231, p. 389 (Dordrecht: Kluwer)
\ref Collins C. A., Guzzo L., Nichol R. C. \& Lumsden S. L., 1995, MNRAS,
274, 1071
\ref Colless M., 1998, in {\it Wide Field Surveys in Cosmology} p. 77 (Paris:
Editions Frontieres)
\ref Colless M. et al, 2001, MNRAS, submitted 
\ref Crone M. M., \& Geller M., 1995, AJ, 110, 21
\ref Cross N. D., et al. 2001, MNRAS, in press
\ref Dalton G. B., Efstathiou G., Maddox S. J., \& Sutherland W. J., 1992,
ApJ, 390, L1
\ref Dalton G. B., Maddox S. J., Sutherland W. J., \& Efstathiou G., 1997,
MNRAS 289, 263
\ref Danese L., de Zotti G., \& di Tullio G., 1980, A\&A, 82, 322
\ref De Grandi S. et al 1999, ApJ, 514, 148
\ref Dubinski J., 1998, ApJ, 502, 141
\ref Girardi M., Biviano A., Giuricin G., Mardirossian F., \& Mezzetti M.,
1993, ApJ, 404, 38
\ref Katgert P., et al. 1996, A\&A, 310, 8
\ref Loveday J., Peterson B. A., Efstathiou G., Maddox S. J. 1992, ApJ, 390, 338
\ref Lucey J. R., 1988, MNRAS, 204, 33
\ref Lumsden S. L., Nichol R. C., Collins C. A., \& Guzzo L. 1992, MNRAS, 258, 1
\ref Maddox S. J., Efstathiou G., Sutherland W. J., \& Loveday J., 1990a, MNRAS, 243, 692
\ref Maddox S. J., Efstathiou G., \& Sutherland W. J., 1990b, MNRAS, 246, 433
\ref Maddox S. J., Sutherland W. J., Efstathiou G., Loveday, J. \& Peterson B.
A. 1990c, MNRAS, 247, 1p
\ref Maddox S. J., et al. 1998 in {\it Large Scale Structure: Tracks and Traces} eds.
V. M{\"u}ller, S. G. Gottl{\"o}ber, J. P. M{\"u}cket , and J. Wambsganss, p. 91
(Singapore: World Scientific)
\ref Mazure A. et al. 1996, A\&A, 310, 31
\ref Postman M., Huchra J. P., Geller M., \& Henry J. P., 1985, AJ, 90, 1400
\ref Schuecker P. et al. 2001, A\&A, 368, 86
\ref Scott E. L., 1956, AJ, 61, 190
\ref Shectman S. A., 1985, ApJS, 57, 77
\ref Struble M. F., \& Rood H. J., 1999, ApJS, 125, 35
\ref Sutherland W., 1988, MNRAS, 234, 159
\ref Sutherland, W., \& Efstathiou G., 1991, MNRAS, 248, 159
\ref van Haarlem M. P., Frenk C. S. \& White S. D. M., 1997, MNRAS, 287, 817
\ref Yahil A., \& Vidal N. V., 1977, ApJ, 214, 347
\ref Zabludoff A. I., Geller M. J., \& Huchra J. P., 1990, ApJS, 74, 1
\ref Zabludoff A. I., Geller M. J., Huchra J. P. \& Ramella M., 1993, AJ, 106, 1301
\ref Zucca E., et al. 1997, A\&A, 326, 477
\end{document}